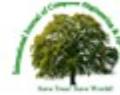

# CONGESTION CONTROL TECHNIQUE USING INTELLIGENT TRAFFIC AND VANET

**Sunil Kumar Singh[1], Rajesh Duvvuru [2], and Saurabh Singh Thakur [3]**

*[1]Department of Computer Science & Engieering*

*National Institute of Technology, Jamshedpur,*

*Jharkhand, India*

**ABSTRACT:**

Road traffic jams is a most important problem in nearly all cities around the world, especially in developing regions resulting in enormous delays, increased fuel wastage and monetary losses. In this paper, we have obtained an in-sight idea of simulating real world scenario of a critical region where traffic congestion is very high. As it is not easy to set up and implement such a complicated system in real world before knowing the impact of all parameters used in Vehicle Ad hoc Network (VANET), a small real world area i.e. Jamshedpur Regal area was taken into consideration, for studying the impact of mobility in the VANET. Traffic movement has been deployed across the area under consideration using one of the realistic vehicular mobility models. The behavior of this network was simulated using SUMO and NS2 to study the impact of traffic and traffic lights at the intersection on packet transmission over Vehicle to Vehicle (V2V) communication using Ad hoc On Demand Vector (AODV) routing protocol and IEEE 802.11 standard. From the traffic simulation results, it is proved that congestion problem for the given sample map with traffic lights and predefined flow can be solved. There was negligible congestion at lanes when traffic logic was changed in accordance with allowing flow with high traffic overload to have high priority than those with low load.

## [I] INTRODUCTION

According to the World Health Organization, road traffic injuries caused an estimated 1.24 million deaths worldwide in the year 2012 [1]. According to an online article published, "India's record in deaths has touched a new low, as toll rose to at least 14 deaths per hour in 2011 against 13 the previous year". While trucks/lorries and two-wheelers were responsible for over 40% deaths, the rush during afternoon and evening hours were the most fatal phases. India leads the world in road deaths. In addition to this, some of the common problems to tackle with are the "Miles of Traffic Jam" on highway and the "Search for best Parking Lot" in an unknown city. For all the above mentioned reasons, the Government and Automotive Industries today pay lot of attention towards traffic management and regulation of a smooth traffic. They are investing many resources to slow down the adverse effect of transportation on environment, thereby increasing traffic efficiency and road safety. The advancements in technology, in the areas of Information and Communications, have opened a new range of possibilities.

VANET is characterized as a special class of Mobile Ad hoc Networks (MANETs) which consists of number of vehicles with the capability of communicating with each other without a fixed infrastructure.

The goal of VANET research is to develop a vehicular communication system to enable 'quick' and 'cost-efficient' transmission of data for the benefit of passenger's safety and comfort. Due to the expensive cost of deploying and complexity of implementing such a system in real world, research in VANET relies on simulation. However, the simulation depends on the mobility model that represents the movement pattern of mobile users including its location, velocity and acceleration over time. A mobility model needs to be a Realistic Mobility Model that considers the characteristics of the real world scenario either by taking a real world MAP obtained from Topologically Integrated Geographic Encoding and Referencing(TIGER)[14] database from U.S. Census Bureau or by taking maps from Open Street Map (OSM) [2] into consideration to simulate a realistic network. In V2V communication or Inter-Vehicle communication, Vehicles are able to communicate with other ongoing vehicles on their path. In this scenario, it is not known in advance when it is possible to meet another vehicle to which the communication is feasible. In vehicle to fixed infrastructure communication (V2I), vehicles are able to communicate with Road Side Unit (RSUs) or access points. In Inter-Road Side Communication it is possible to know the communicating parties in advance as RSUs are placed at a fixed distance from each other. Therefore, the main difference between V2V and V2I is the coverage area [3, 4].

### 1.1 Problem Statement

In this project, we consider the problem of finding the best integrated traffic signal phase plan at a single intersection.

### 1.2 Scope of the Study

## [II] LITERATURE REVIEW

One of the most promising areas is the study of the communication among vehicles and Road Side Units International Journal of Wireless & Mobile Networks, which lead to the emergence of Vehicular Network or Vehicular Ad-hoc Network (VANET) into picture [5]. Research is being carried out in the field of VANET such as analyzing data dissemination in VANETs, Identifying and studying routing protocols in VANET in terms of highest delivery ratio and lowest end-to-end delay etc. The issues of Security and Privacy also demands great attention. The study of Mobility Models and their realistic vehicular model deployment is a challenging task [6]. Random way Point(RWP)[7] is an earlier mobility model widely used in VANET with different routing protocols like AODV, DSR and DSDV[8] in which nodes move freely in a predefined area but without considering any obstacle in that area. However, in a VANET environment vehicles are typically restricted by streets, traffic light and obstacles. Kim J. et. al. [9] presents the mobility model in which vehicles know from the start their initial and final points. The routing track is then chosen considering the social relation between the vehicles and also the destination point. This means that vehicles move only between the initial and final point on path chosen by the social relation strength between the vehicles.

GrooveSim was the first tool for forecasting vehicular traffic flow and evaluating vehicular performance. It gives a traffic simulator environment which is easy to use for generating real traffic scenario for evaluation. But it fails to include network simulator as it was unable to create traces for network. David R. Choffnes et al. [10] proposed a mobility model named STRAW (Street Random Waypoint). This model has taken real map data of US cities and considered the node (vehicle) movement on streets based on this map. This model also has the functionality to simplify the traffic congestion by controlling the vehicular mobility. But still it lacks overtaking criteria that causes convey effect in street as it considered random method which is not realistic. Behrisch, M. et al.[11] describes a realistic tool MOVE for International Journal of Wireless & Mobile Networks  generating realistic vehicular mobility model. It is built on top of an open source micro-traffic simulator SUMO[18] and its output is a realistic mobility model that can immediately be used by popular network simulators such as Ns-2 and Qualnet. Bettstetter et al. [16] gave a Random Direction Model which introduces a stop turn- and-go behavior which can mimic the vehicle behavior at the intersections. Saha et al. [17] proposed a macro mobility model based on TIGER map database.

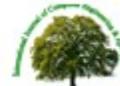

## [III] PROPOSED WORK AND METHODOLOGY

The basic idea proposed in this model is that RSU attached to traffic light will monitor traffic from all directions and various flows using its multi-channel wireless communication unit. RSU will have information regarding flow. This will help RSU to dynamically change traffic light logic to allow flows with heavy load more time with green light than red. This will be compensated by vehicles on low load lanes to spend more time waiting.

The above idea has been proved using a sample map (Jamshedpur Regal area) and a predefined flow between different lanes. This causes congestion in two of the lanes. Later, on changing traffic light logic and giving more priority to flow that cause congestion [flow from lane id= 60263428#5" to lane id="-60260578# and from lane id="117479914#1" to lane id="-60260578#1"]. As will be seen, there is no congestion after new logic is applied. The flows with congestion were prioritized and had more time with green signal to allow vehicles with same flows move and clear congestion.

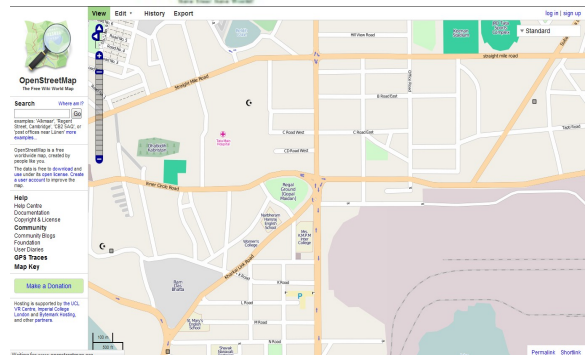

**Fig: 1. OSM map of Jamshedpur Regal area to be imported**

### 3.1. Static-Traffic Logic

The simulation is loaded using SUMO-GUI which is the graphical version of SUMO simulator. The map of Jamshedpur area which is to be simulated is exported from OSM and edited using JOSM. JOSM helps installing traffic lights using it rich user interface. The map os then saved as .osm file and converted to .net.xml file using NETCONVERT tool provided by SUMO. The traffic logic is a part of script file.net.xml which is predefined in map configuration file. This flow.xml with .net.xml file is used to create vehicles traffic scenario using MOVE [13]. The output is rou.xml and rou.alt.xml file which define routes vehicles will take at different time instants of simulation. Start and end time of simulation is provided.

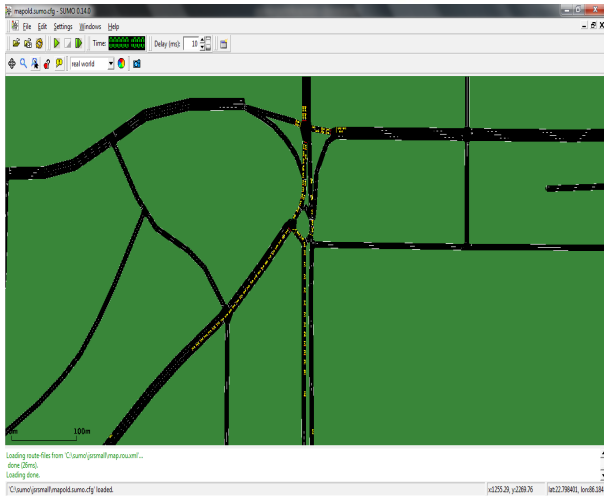

**Fig: 2. Simulation at the end of t=1000ms with static-traffic logic**

## 3.2. Dynamic-Traffic Logic

Since there was traffic congestion problem in previous scenario, we modified the traffic light logic governing state transitions of traffic light to allow vehicles in congested area to pass. When the simulation is run with new traffic logic, the result is an efficient traffic flow in the same scenario of flow.xml, net.xml.

The main objectives towards developing the successful technique are as follows:

1.  Develop a heterogeneous traffic flow model with respect to Traffic Simulator SUMO.

2.  Connect SUMO externally to extended network simulator NS2 using TCP port.

3.  Develop a program for traffic signals' state which can be requested by network simulator to traffic simulator via TraCI interface during online simulation.

4.  Extract various parameters from detectors using TraCI commands.

5.  Verify the traffic congestion control technique

```
<tlLogic id="1274361397" type="static" programID="0"
offset="0">
        <phase duration="31" state="GG"/>
        <phase duration="9" state="yy"/>
    </tlLogic>
    <tlLogic id="1284510665" type="static" programID="0"
offset="0">
        <phase duration="31" state="GGGrrr"/>
        <phase duration="9" state="yyyyrrr"/>
        <phase duration="31" state="rrrgggg"/>
        <phase duration="9" state="rrryyyy"/>
    </tlLogic>
    <tlLogic id="1284510687" type="static" programID="0"
offset="0">
        <phase duration="31" state="GGGGg"/>
        <phase duration="9" state="yyyy"/>
    </tlLogic>
    <tlLogic id="1274361418" type="static" programID="0"
offset="0">
        <phase duration="31" state="rrrrGGGggg"/>
        <phase duration="9" state="rrryyyyggg"/>
        <phase duration="6" state="rrrrrrrggg"/>
        <phase duration="31" state="rrrrrrgyyy"/>
        <phase duration="31" state="gggggrrgrrr"/>
        <phase duration="9" state="yyyyrrrrrr"/>
    </tlLogic>
    <tlLogic id="748825800" type="static" programID="0"
offset="0">
        <phase duration="31" state="gggrrr"/>
        <phase duration="9" state="yyyrrr"/>
        <phase duration="31" state="rrrGGG"/>
        <phase duration="9" state="rrryyy"/>
    </tlLogic>
```

**Fig: 3. Static-Traffic Logic**

```
<tlLogic id="1274361397" type="static" programID="0" offset="0">
        <phase duration="31" state="GG"/>
        <phase duration="9" state="yy"/>
    </tlLogic>

    <tlLogic id="1284510665" type="static" programID="0"
offset="0">
        <phase duration="31" state="GGGrrr"/>
        <phase duration="9" state="yyyyrrr"/>
        <phase duration="31" state="rrrgggg"/>
        <phase duration="9" state="rrryyyy"/>
    </tlLogic>
    <tlLogic id="1284510687" type="static" programID="0"
offset="0">
        <phase duration="31" state="gggg"/>
        <phase duration="9" state="yyyy"/>
    </tlLogic>
    <tlLogic id="1274361418" type="static" programID="0"
offset="0">
        <phase duration="31" state="rrrrGGGrrr"/>
        <phase duration="9" state="rrryyyyrrr"/>
        <phase duration="6" state="rrrrGGGrrr"/>
        <phase duration="9" state="rrrrGGGrrr"/>
        <phase duration="31" state="gggggrrGGG"/>
        <phase duration="9" state="yyyyrrrggg"/>
    </tlLogic>
    <tlLogic id="748825800" type="static" programID="0"
offset="0">
        <phase duration="31" state="GGGrrr"/>
        <phase duration="9" state="yyyrrr"/>
        <phase duration="31" state="rrrggg"/>
        <phase duration="9" state="GGGrrr"/>
    </tlLogic>
```

**Fig: 4. Dynamic-Traffic Logic**

SUMO is fed with the input files to prepare the network for the simulation. Two of the files contain the geometric network information, which will be converted into node and link information in SUMO and are named with extensions .nod.xml and .edg.xml respectively. The file containing the vehicle types to be used, traffic demand and respective route information of each vehicle will be named with extension .rou.xml. In addition, file with extension .con.xml is included for specifying the allowed traffic movements in the network. The file with extension .det.xml can be used to specify the location of detectors.

The nodes, edges and connections files are combined using the application NETCONVERT in the simulator to develop the geometric network. The network file combined with routes file and detectors file is used to develop the configuration file with extension .sumo.cfg which when given as input to SUMO or SUMO-GUI application generates the simulation. During the run-time of simulation, SUMO is connected via TraCI interface [11] to network simulator which requests the parameters from SUMO and sends commands to change the state of traffic signals after every simulation time step.

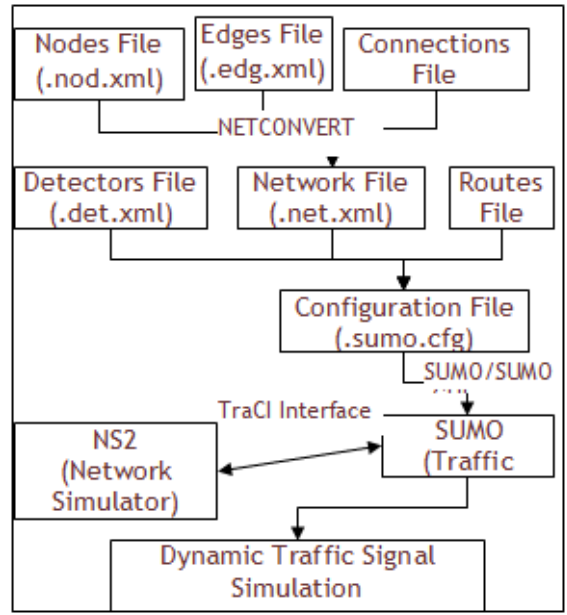

**Fig: 3. Flow Chart of Input Files for Dynamic Traffic Simulation Process**

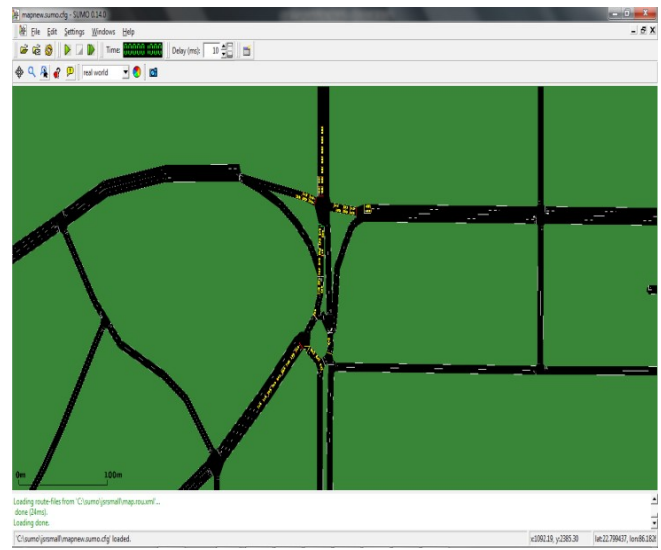

**Fig: 4. SUMO-GUI at the end of t=1000ms with new traffic logic**

## [IV] SIMULATION RESULTS AND DISCUSSION

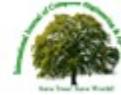

MOVE consists of two components, the Map Editor and the Vehicle Movement Editor. The Map Editor used to incorporate real world existing map such as publically available TIGER (Topologically Integrated Geographic Encoding and Referencing) database from U.S. Census Bureau allows users to rapidly generate realistic VANET mobility models in two aspects:-by interfacing with real world map databases such as [14], MOVE allows user to conveniently incorporate realistic road maps into the simulation. The SUMO simulator is connected to network simulator NS2[15] using TraCI interface [12] and study the impact of driver's choice with traffic lights at the intersection on packet transmission over V2V communication using AODV routing protocol and IEEE 802.11 standard. Based on these simulations results our dynamic traffic light will change. Initially, the green signal is given to west-east direction. Network-simulator requests for the values of the various parameters after every simulation time step through TraCI Interface. The traffic-simulator sends the parameters and listens for the incoming requests from network-simulator. The network-simulator requests the traffic-simulator to change the state of various parameters to continue the simulation. After simulation ns2 script we get two files trace file and nam file AWK script is used to start the TCP port connection, request parameters, exchange requests and responses. The impact of realistic vehicular mobility using various tools on the performance of ad-hoc routing protocols has been evaluated in this section. The driver route choice behavior with traffic lights at the intersections has been simulated for a real world scenario. In this, all possible routes from the source to destination are defined and the driver needs to decide about which route is to be taken from among all possible routes at any intersection. The presence of traffic lights at the intersection regulates the smooth movement of vehicles in different directions and causes clustering effect by forcing the vehicles to stop at intersection when the signal is red. Therefore, the node density at the intersection increased which improves the network connectivity among the peers at intersection, but the improved connectivity deteriorates the packet delivery ratio.

Our simulation concentrates on selecting the probability of choosing a route at the intersection. This probability directly determines the number of vehicles on a particular route. The data in terms of packets are transmitted to facilitate communication among vehicles. In order to study the behavior of communication, the parameters like packet delivery fraction, average duration of packets/sec and average duration of bits/sec has been considered which are discussed in the subsequent sections.

## 4.1. Packet Delivery Fraction

It is the ratio of number of packets delivered to numbers of packets generated by traffic generator. Higher the PDF, greater is the efficiency of network.

It can be calculated as:

Packet delivery fraction = (Number of packets received) / (Number of packets sent)

| Number of nodes | PDF |
|---|---|
| 10 | 86.26% |
| 20 | 93.38% |
| 30 | 89.24% |
| 40 | 88.58% |
| 50 | 90.88% |

**Table: 1. PDF on various numbers of nodes**

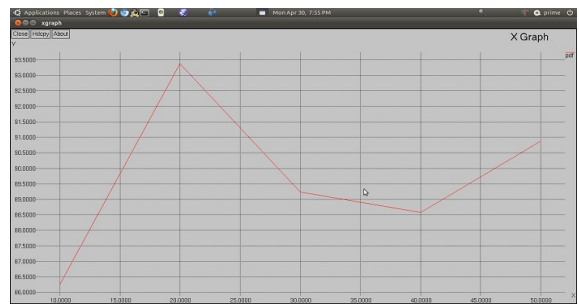

**Fig: 5. Packet Delivery Fraction**

It can be observed from table and graph that PDF increases on increasing of nodes or vehicles. So for large amount of traffic it will performed better.

## 4.2 Average Duration of Packets/second

From the below table and graph we can see the average duration of packets per second. It shows that when number of nodes are increasing the packets duration time is also increasing.

| Number of Nodes | AD of Packets/sec |
|---|---|
| 10 | 2.44 |
| 20 | 2.43 |
| 30 | 7.01 |
| 40 | 3.62 |
| 50 | 4.04 |

**Table: 2. Average Duration of Packets/second**

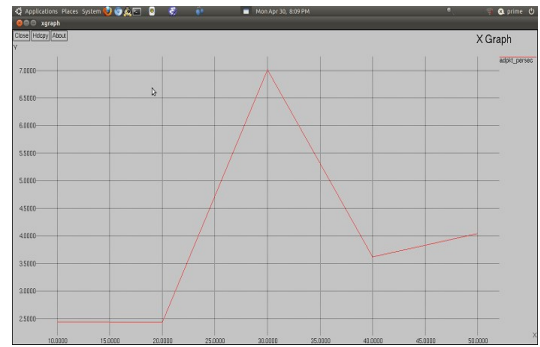

**Fig: 6. Average Duration of Packets/second**

## [V] CONCLUSION

In this paper, we have developed a congestion control technique suitable for Indian heterogeneous traffic. As an extension, we aim at the calibration and validation of our model by comparing the results with various field data. We present a local congestion problem and solution that coordinates traffic signal behavior within a small area and can locally prevent congestion collapse sustaining time variant traffic bursts. Based on a simulation based analysis on simple network topologies, we show that our local congestion problem can enhance road capacity and prevent congestion collapse in localized settings. Traffic movement has been deployed across the area under consideration using one of the realistic vehicular mobility models. From the traffic simulation results, it is proved that congestion problem for the given sample map with traffic lights and predefined flow can be solved. There was negligible congestion at lanes when traffic logic was changed in accordance with allowing flow with high traffic overload to have high priority than those with low load. Both at traffic level and network level the model worked out really well. The idea proposed in this paper can used to solve real world congestion problems at traffic intersections.

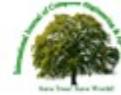